\documentclass{article}
\usepackage{graphicx}
\usepackage{cite}
\begin{document}

\title{Generalized Uncertainty Principle and the Ramsauer-Townsend Effect}

\author{Javad Vahedi,$^{a,}$\thanks{j.vahedi@gmail.com}\quad
        Kourosh Nozari,$^{a,}$\thanks{knozari@umz.ac.ir} \, and
        Pouria Pedram$^{b,}$\thanks{p.pedram@srbiau.ac.ir}\\
        {\small $^{a}$Department of Physics, Islamic Azad University, Sari
        Branch, Sari, Iran}\\
        {\small $^{b}$Department of Physics, Science and Research Branch,
        Islamic Azad University, Tehran, Iran}}

\date{\today}
\maketitle \baselineskip 24pt

\begin{abstract}
The scattering cross section of electrons in noble gas atoms
exhibits a minimum value at electron energies of approximately
$1$eV. This is the Ramsauer-Townsend effect. In this letter, we
study the Ramsauer-Townsend effect in the framework of the
Generalized Uncertainty Principle.
\end{abstract}


\textit{Pacs}: {04.60.-m}\\

\section{Introduction}
Various approaches to quantum gravity, such as string theory and
loop quantum gravity as well as black hole physics, predict a
minimum measurable length of the order of the Planck length,
$\ell_{p}=\sqrt{\frac{G\hbar}{c^{3}}}\sim10^{-35}m$. In the presence
of this minimal observable length, the standard Heisenberg
Uncertainty Principle attains an important modification leading to
the so-called Generalized Uncertainty Principle (GUP). As a result,
corresponding commutation relations between position and momenta are
generalized too \cite{1}. In recent years a lot of attention has
been attracted to extend the fundamental problems of physics in this
framework (see for instance
\cite{21,22,23,24,25,26,27,28,29,30,31,32,33,34,35,36,37,38,39,40,p,PRD,PhysA,PLB}).
Since in the GUP framework one cannot probe distances smaller than
the minimum measurable length at finite time, we expect it modifies
the Hamiltonian of systems too. Recently it has been shown that the
GUP affects Lamb shift, Landau levels, reflection and transmission
coefficients of a potential step and potential barrier \cite{9}. In
addition, they speculated on the possibility of extracting
measurable predictions of GUP in the future experiments. In this
work we will follow the procedure introduced in the Ref. \cite{9},
but we are going to address the effect of GUP on the
Ramsauer-Townsend (RT) effect. The RT effect can be observed as long
as the scattering does not become inelastic by excitation of the
first excited state of the atom. This condition is best fulfilled by
the closed shell noble gas atoms. Physically, the RT effect may be
thought of as a diffraction of the electron around the rare-gas
atom, in which the wave function inside the atom is distorted in
such a way that it fits on smoothly to an undistorted wave function
outside. The effect is analogous to the perfect transmission found
at particular energies in one-dimensional scattering from a square
well. The one-dimensional treatment of scattering from a square well
and also three-dimensional treatment using the partial waves
analysis can be found in \cite{14}. We generalize the
one-dimensional treatment of the scattering from a square well to
the GUP framework. We also address the condition for interference in
the Fabry-Perot interferometer in the framework of GUP.

\section{A Generalized Uncertainty Principle}
Quantum mechanics with modification of the usual canonical
commutation relations has been investigated intensively in the last
few years (see \cite{PRD} and references therein). Such works which
are motivated by several independent streamlines of investigations
in string theory and quantum gravity, suggest the existence of a
finite lower bound to the possible resolution $\Delta X$ of
spacetime points. The following deformed commutation relation has
attracted much attention in recent years \cite{1}
\begin{equation}
[X, P]=i\hbar(1+\beta P^2),
\label{eq1}
\end{equation}
and it was shown that it implies the existence of a minimal
resolution length $\Delta X=\sqrt{\langle X^2 \rangle -\langle X
\rangle^2}\ge\hbar\sqrt\beta$. This means that there is no
possibility to measure coordinate $X$ with accuracy smaller than
$\hbar\sqrt\beta$. Since in the context of the string theory the
minimum observable distance is the string length, we conclude that
$\sqrt{\beta}$ is proportional to this length. If we set $\beta=0$,
the usual Heisenberg algebra is recovered. The use of the deformed
commutation relation (\ref{eq1}) brings new difficulties in solving
the quantum problems. A part of difficulties is related to the break
down of the notion of locality and position space representation in
this framework \cite{1}. The above commutation relation results in
the following uncertainty relation:
\begin{eqnarray}
 \Delta X \Delta P \geq \frac{\hbar}{2}
\left( 1 +\beta (\Delta P)^2 +\gamma \right),
\label{eq2}
\end{eqnarray}
where $\beta$ is the GUP parameter and $\gamma$ is a positive
constant that depends on the expectation value of the momentum
operator. In fact, we have $\beta=\beta_0/(M_{Pl} c)^2$ where
$M_{Pl}$ is the Planck mass and $\beta_0$ is of the order of unity.
We expect that these quantities are only relevant in the domain of
the Planck energy $M_{Pl} c^2\sim 10^{19}$GeV. Therefore, in the low
energy regime, the parameters $\beta$ and $\gamma$ are irrelevant
and one recovers the well-known Heisenberg uncertainty principle.
These parameters, in principle, can be obtained from the underlying
quantum gravity theory such as string theory. Moreover, the
comparison between Eqs.~(\ref{eq1}) and (\ref{eq2}) shows that
$\gamma=\beta\langle P\rangle^2$. Now, let us define \cite{9}
\begin{eqnarray}
\left\{
\begin{array}{ll}
X = x,\\\\ P = p \left( 1 + \frac{1}{3}\beta\, p^2 \right),
\end{array}
\right.
\label{eq4}
\end{eqnarray}
where $x$ and $p$ obey the canonical commutation relations
$[x,p]=i\hbar$. One can check that using Eq.~(\ref{eq4}),
Eq.~(\ref{eq1}) is satisfied up to ${\cal{O}}(\beta)$. Also, from
the above equation we can interpret $p$ as the momentum operator at
low energies ($p=-i\hbar \partial/\partial{x}$) and $P$ as the
momentum operator at high energies. Now, consider the following form
of the Hamiltonian:
\begin{eqnarray}
H=\frac{P^2}{2m} + V(x),
\label{eq5}
\end{eqnarray}
which using Eq.~(\ref{eq4}) can be written as
\begin{eqnarray}
H=H_0+\beta H_1+{\cal{O}}(\beta^2),
\label{eq6}
\end{eqnarray}
where $H_0=\frac{\displaystyle p^2}{\displaystyle2m} + V(x)$ and
$H_1=\frac{\displaystyle p^4}{\displaystyle3m}$.

In the quantum domain, this Hamiltonian results in the following
generalized Schr\"odinger equation in the quasi-position
representation
\begin{eqnarray}
-\frac{\hbar^2}{2m}\frac{\partial^2\psi(x)}{\partial
x^2}+\beta\frac{\hbar^{4}}{3m}\frac{\partial^{4}\psi(x)}{\partial
x^{4}} +V(x)\psi(x)=E\psi(x),
\label{eq7}
\end{eqnarray}
where the second term in the left side is due to the generalized
commutation relation (\ref{eq1}). This equation is a fourth-order
differential equation which in principle admits four independent
solutions. Therefore, solving this equation in $x$ space and
separating the physical solutions is not an easy task. With these
preliminaries, in the next section we solve equation (\ref{eq7}) for
a quantum well to address the RT effect and the Fabry-Perot
interferometer resonance condition in the presence of the minimal
observable length.

\section{The Ramsauer-Townsend effect with GUP}
\begin{figure}
\centering
\includegraphics[width=8cm]{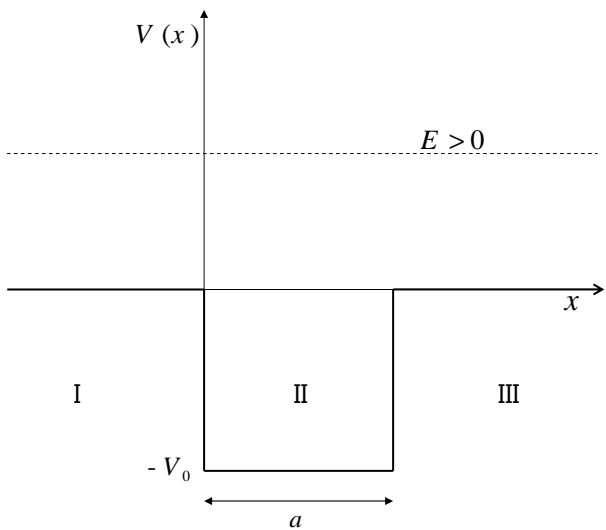}
\caption{ The geometry of a quantum well.}\label{fig1}
\end{figure}

We choose the following geometry of the quantum well (see
Fig.~\ref{fig1})
\begin{equation}
V(x)=\left\{\begin{array}{ll} -V_{0}&\quad \quad0< x < a,\\ \\
\quad0&\quad\quad{\rm elsewhere},\end{array}\right. \label{eq8}
\end{equation}
where $V_{0}$ is a positive constant and we assume $E>0$.

\begin{figure}
\centering
\includegraphics[width=8cm]{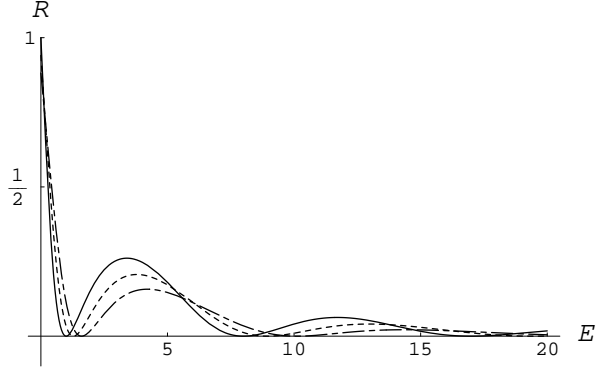}
\caption{The reflection coefficient $R$ versus the energy $E$ for
$\beta=0$ (solid line), $\beta=0.005$ (dashed line), $\beta=0.01$
(dot-dashed line), $V_0=8$, $a=\pi$, and $\hbar=2m=1$.}\label{fig2}
\end{figure}

The eigenfunctions of a particle in this potential well satisfy the
generalized Schr\"{o}dinger equation (\ref{eq7}). We need to find
the solutions in three different regions which are indicated in
Fig.~(\ref{fig1}). To proceed further, we rewrite Eq.~(\ref{eq7}) in
these regions separately as
\begin{equation}
d^{2}\psi(x)+q^{2}\psi(x)-\ell_{p}^{2}d^{4}\psi(x)=0,  \label{eq12}
\end{equation}
for $0<x<a$, and
\begin{equation}
d^{2}\psi(x)+k^{2}\psi(x)-\ell_{p}^{2}d^{4}\psi(x)=0,  \label{eq13}
\end{equation}
elsewhere, where by definition
$d^{n}\equiv\frac{\partial^{n}}{\partial x^{n}}$,
$k=\sqrt{\frac{2mE}{\hbar^{2}}}$,
$q=\sqrt{\frac{2m(E+V_{0})}{\hbar^{2}}}$, and
$\ell_{p}=\hbar\sqrt{\frac{2\beta}{3}}$. As state before, the above
equations are forth-order differential equations which in general
admit four independent solutions. However, some solutions would be
unphysical which should be removed upon imposing the boundary
conditions. As it is shown in Ref.~\cite{p}, alternatively, we can
find the equivalent physical solutions by adding the following
constraint: the physical solutions should also satisfy the ordinary
Schr\"{o}dinger equation but with different eigenenergy. In fact,
for the cases of a free particle and a particle in a box, this
additional condition prevents us from doing equivalent but lengthy
calculations \cite{p}. Therefore, we demand that the eigenfunctions
also satisfy the following second-order differential equations:
\begin{equation}
d^{2}\psi(x)+q'^{2}\psi(x)=0,  \label{eq12a}
\end{equation}
for $0<x<a$, and
\begin{equation}
d^{2}\psi(x)+k'^{2}\psi(x)=0,  \label{eq13a}
\end{equation}
elsewhere, where $k'=\sqrt{\frac{2mE'}{\hbar^{2}}}$ and
$q'=\sqrt{\frac{2m(E'+V_{0})}{\hbar^{2}}}$. The solutions of
Eqs.~(\ref{eq12a}) and (\ref{eq13a}) in regions $\mathrm{I}$,
$\mathrm{II}$ and $\mathrm{III}$ are
\begin{eqnarray}\left\{
\begin{array}{l}
\psi_{\mathrm{I}}=e^{ik'x}+Ae^{-ik'x},\\\\
\psi_{\mathrm{II}}=Be^{iq'x}+Ce^{-iq'x},\\\\
\psi_{\mathrm{III}}=De^{ik'x},
\end{array}\right.
\end{eqnarray}
respectively. These solutions should also satisfy Eqs.~(\ref{eq12})
and (\ref{eq13}) which result in
\begin{equation}\label{kq}
k^2=k'^2+\ell_{p}^{2}k'^4,\hspace{2cm} q^2=q'^2+\ell_{p}^{2}q'^4.
\end{equation}
These solutions are similar to the solutions of the ordinary quantum
mechanics but with modified wavenumbers. Now the boundary conditions
are the continuity of the wave functions and their first derivatives
at the boundaries. The resulting equations can be solved
analytically to obtain the coefficients $A$, $B$, $C$, and $D$. For
our purposes, the solution for $A$ is as follows
\begin{equation}
A=\frac{(k'^{2}-q'^{2})\sin(q'a)}{(k'^2+q'^2)\sin(q'a)+2ik'q'\cos(q'a)}\,.
\label{eq30}
\end{equation}
So the reflection coefficient is given by
\begin{eqnarray}
R_a(k',q')&\equiv&|A|^{2},\nonumber \\ &=&
\frac{(k'^{2}-q'^{2})^2\sin^2(q'a)}{(k'^2+q'^2)^{2}\sin^2(q'a)+4k'^2q'^2\cos^2(q'a)}.
\label{eq31}
\end{eqnarray}
Because of the smallness of the Planck length, we can obtain
$k'\simeq k\left(1-\frac{1}{2}\ell_{p}^{2}k^2\right)$ and $q'\simeq
q\left(1-\frac{1}{2}\ell_{p}^{2}q^2\right)$ from Eq.~(\ref{kq}) and
write the reflection coefficient in terms of the physical
wavenumbers
\begin{equation}
R_a(k,q)=\frac{(k^{2}-q^{2})^2\left[1-2\ell_{p}^{2}(k^2+q^2)\right]
\sin^2\left[q\left(1-\frac{1}{2}\ell_{p}^{2}q^2\right)a\right]}{\left[(k^2+q^2)^{2}-
2\ell_{p}^{2}(k^4+q^4)\right]\sin^2\left[q\left(1-\frac{1}{2}\ell_{p}^{2}q^2\right)a\right]
+4k^2q^2
\left[1-\ell_{p}^{2}(k^2+q^2)\right]\cos^2\left[q\left(1-\frac{1}{2}\ell_{p}^{2}q^2\right)a\right]}.\label{wide}
\end{equation}
Figure \ref{fig2} shows the variation of the reflection coefficient
versus the energy. This figure compares also the ordinary quantum
mechanical result with corresponding result in the presence of a
minimal observable length.

\begin{figure}
\centering
\includegraphics[width=8cm]{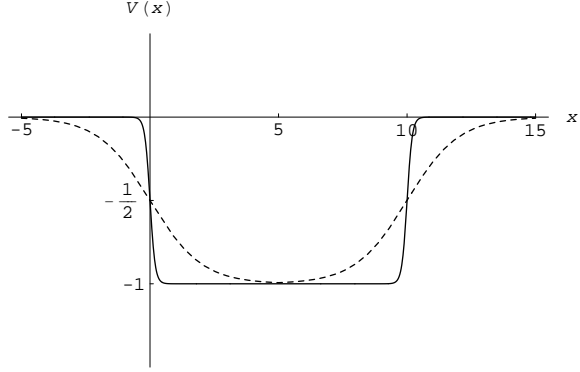}
\caption{The Woods-Saxon potential well for $L = 10$ with $\alpha
=10$ (solid line) and $\alpha = 1$ (dashed line).}\label{fig3}
\end{figure}

At this point, it is worth to mention that the rectangular potential
well is an idealization, and it would be desirable to evaluate the
admitted deviations of a real potential from this ideal one. Indeed,
in reality, the sharp edges are changed to the smoothed out edges. A
proper candidate for this case is the Woods-Saxon potential which
has the following functional form \cite{r1}:
\begin{equation}
V(x)=-V_0\left[\frac{\theta(-x+L/2)}{1+e^{-\alpha
x}}+\frac{\theta(x-L/2)}{1+e^{\alpha (x-L/2)}}\right],
\end{equation}
where $\alpha$ and $L$ are real and positive, and $\theta(x)$ is the
Heaviside step function. For $\alpha L\gg1$ this potential closely
resembles a rectangular well with smooth edges and size $L$
(Fig.~\ref{fig3}). This potential is exactly solvable in
relativistic and non-relativistic cases and the solutions can be
written in terms of the hypergeometric functions \cite{r1,r2,r3}.
Since these solutions smoothly converge to the plane wave solutions
for $\alpha L\gg1$, we expect that the GUP-corrected reflection
coefficient of this system also continuously tends to
Eq.~(\ref{wide}) at this limit. However, for this case the wave
function cannot satisfy both the ordinary and GUP-corrected
Schr\"odinger equations simultaneously. This is due to the fact that
the potential is not constant inside the well. This problem needs
further investigation and we are going to study it in a separate
program.

For the particular case where
$\sin[q\left(1-\frac{1}{2}\ell_{p}^{2}q^2\right)a]=0$, there is no
reflection, that is $R=0$ and therefore we will have maximum
transmission. This is the Ramsauer-Townsend effect. In this case
\begin{equation}
q\left(1-\frac{1}{2}\ell_{p}^{2}q^2\right)=\frac{n\pi}{a}.
\label{eq32}
\end{equation}
In ordinary quantum mechanics this effect occurs at those
wavenumbers that satisfy the condition
$q_{\mbox{\footnotesize{ord}}}=n\pi/a$. This feature shows that
there is a shift
($\Delta_q=q_{\mbox{\tiny{GUP}}}-q_{\mbox{\footnotesize{ord}}}$) in
the wavenumber of the transmission resonance and this shift itself
is wavenumber-dependent. Up to the first-order in the GUP parameter,
we find
\begin{equation}
\Delta_q\simeq\frac{1}{2}\ell_{p}^{2}\left(\frac{n\pi}{a}\right)^3.
\end{equation}
We also note that in ordinary quantum mechanical description, the
condition for resonance is
$\lambda_{\mbox{\footnotesize{ord}}}=2\pi/q=2a/n$ which is the same
condition as in Fabry-Perot interferometer. In the presence of the
minimal observable length, this condition modifies as follows
\begin{equation}
\lambda'=\frac{2\pi}{q'}\simeq\frac{2\pi}{q}\left(1+\frac{1}{2}\ell_{p}^{2}q^{2}\right)=
\lambda_{\mathrm{ord}}\left(1+\frac{1}{2}\ell_{p}^{2}q^{2}\right).
\label{eq34}
\end{equation}
Therefore, in the presence of the minimal length, the condition for
interference in Fabry-Perot interferometer will change too.
Amazingly, this change is itself wavelength-dependent.

Up to this point, we have addressed the RT effect and the
Fabry-Perot interferometer resonance condition in the GUP framework.
To complete our treatment of this interesting quantum mechanical
problem, let us consider the negative energy case $-V_0<E<0$ which
results in the quantized energy spectrum. In this case,
Eqs.~(\ref{eq12a}) and (\ref{eq13a}) cast into the following
equations:
\begin{equation}
d^{2}\psi(x)+q'^{2}\psi(x)=0,  \label{eq12b}
\end{equation}
for $0<x<a$, and
\begin{equation}
d^{2}\psi(x)-\kappa'^{2}\psi(x)=0,  \label{eq13b}
\end{equation}
elsewhere, where by definition
$\kappa=\sqrt{\frac{2m|E'|}{\hbar^2}}$ and
$q=\sqrt{\frac{2m(V_0-|E'|)}{\hbar^2}}$. So the solutions are
\begin{eqnarray}\left\{
\begin{array}{l}
\psi_{\mathrm{I}}=Ae^{\kappa'x},\\\\
\psi_{\mathrm{II}}=Be^{iq'x}+Ce^{-iq'x},\\\\
\psi_{\mathrm{II}}=De^{-\kappa'x}.
\end{array}\right.
\end{eqnarray}
If we choose the center of the well as the center of the coordinate
system, it is straightforward to check that the energy eigenvalues
are given by the roots of equations
\begin{eqnarray}
\left\{
\begin{array}{l}
\tan(q'a/2)=\kappa'/q', \\\\
\cot(q'a/2)=-\kappa'/q',
\end{array}\right.
\end{eqnarray}
for even and odd eigenstates, respectively. These eigenvalues can
also be written in terms of physical quantities $k$ and $q$ as
\begin{eqnarray}
\tan\left[q\left(1-\frac{1}{2}\ell_{p}^{2}q^2\right)a/2\right]&=&
\frac{\sqrt{\frac{2mV_0}{\hbar^2}-q^2}\left[1-\frac{1}{2}\ell_{p}^{2}
\left(\frac{2mV_0}{\hbar^2}-q^2\right)\right]}{q\left(1-\frac{1}{2}\ell_{p}^{2}q^2\right)},\\
\cot\left[q\left(1-\frac{1}{2}\ell_{p}^{2}q^2\right)a/2\right]&=&
-\frac{\sqrt{\frac{2mV_0}{\hbar^2}-q^2}\left[1-\frac{1}{2}\ell_{p}^{2}
\left(\frac{2mV_0}{\hbar^2}-q^2\right)\right]}{q\left(1-\frac{1}{2}\ell_{p}^{2}q^2\right)}.
\label{eq030}
\end{eqnarray}
So, for the negative energy case, the energy eigenvalues are the
roots of Eq.~(\ref{eq030}). We note that there is no trace of the RT
effect for $-V_0<E<0$ case.

\section{Conclusions}
The scattering cross section of electrons in noble gas atoms
exhibits a minimum value at electron energies of approximately
$1$eV, an effect of which is called the Ramsauer-Townsend effect. We
studied the RT effect in the presence of the minimal observable
length in the framework of the generalized uncertainty principle. We
have shown that in the presence of the minimal observable length
there is a shift
$\Delta_q=q_{\mathrm{GUP}}-q_{\mathrm{ord}}\simeq\frac{1}{2}\ell_{p}^{2}\left(\frac{n\pi}{a}\right)^3
$ in the wavenumber of the transmission resonance and this shift
itself is wavenumber dependent. This shift also affects the
resonance in the Fabry-Perot interferometer in such a way that this
change is itself wavelength dependent. If in the future experiments
one finds a similar shift in Fabry-Perot interferometer resonance
wavelength, it would be an explicit trace of underlying quantum
gravity scenario. We note also that the RT effect is in principle a
3-dimensional effect that needs a 3-dimensional analysis. However,
corresponding calculations in 3-dimensions are so lengthy and the
essential ingredients and outcomes are the same as presented in this
one-dimensional analysis. Finally, we note that this problem can be
treated with more real potentials such as the Woods-Saxon potential
to have more realistic situation. This potential is exactly solvable
in relativistic and non-relativistic cases and the solutions can be
written in terms of the hypergeometric functions. Since these
solutions smoothly converge to the plane wave solutions in
appropriate limit, we expect that the GUP-corrected reflection
coefficient of this system also continuously tends to our result at
this limit. We are going to treat this more real situation in a
separate research program.


\begin{thebibliography}{11}
\bibitem{1}         A. Kempf, G. Mangano and R.B. Mann, Phys. Rev. D {\bf52} (1995)
1108\\ K. Nozari and A. Etemadi, Phys. Rev. D, \textbf{85}(2012)
104029
\bibitem{21}        D. Amati, M. Ciafaloni, and G. Veneziano, Phys. Lett. B \textbf{216} (1989) 41.
\bibitem{22}        M. Maggiore, Phys. Lett. B \textbf{304} (1993) 65, [arXiv:hep-th/9301067].
\bibitem{23}        M. Maggiore, Phys. Rev. D \textbf{49} (1994) 5182, [arXiv:hep-th/9305163].
\bibitem{24}        M. Maggiore, Phys. Lett. B \textbf{319} (1993) 83, [arXiv:hep-th/9309034].
\bibitem{25}        L. J. Garay, Int. J. Mod. Phys. A \textbf{10} (1995) 145, [arXiv:gr-qc/9403008].
\bibitem{26}        F. Scardigli, Phys. Lett. B \textbf{452} (1999) 39, [arXiv:hep-th/9904025].
\bibitem{27}        S. Hossenfelder, M. Bleicher, S. Hofmann, J. Ruppert, S. Scherer, and H. Stoecker, Phys. Lett. B \textbf{575} (2003) 85,
arXiv:[hep-th/0305262]\\S. Hossenfelder, [arXiv:1203.6191]\\ S.
Hossenfelder, Class. Quantum Grav. \textbf{29}, 115011 (2012)
\bibitem{28}        C. Bambi, and F.R. Urban, Class. Quantum Grav. \textbf{25} (2008) 095006, [arXiv:0709.1965].
\bibitem{29}        K. Nozari and B. Fazlpour, Gen. Rel. Grav. \textbf{38} (2006) 1661, [arXiv:gr-qc/0601092].
\bibitem{30}        K. Nozari, Phys. Lett. B {\bf629} (2005) 41.
\bibitem{31}        K. Nozari and S.H. Mehdipour, Europhys. Lett. {\bf84} (2008) 20008; Chaos, Solitons and Fractals {\bf32} (2007) 1637.
\bibitem{32}        K. Nozari and S. H. Mehdipour, Class. Quantum Gravit. {\bf 25} (2008)
                    175015;  K. Nozari and S. H. Mehdipour, Mod. Phys. Lett. A {\bf
                    20} (2005) 2937.
\bibitem{33}        M. V. Battisti, G. Montani, Phys. Rev. D \textbf{77}  (2008) 023518, [arXiv:0707.2726].
\bibitem{34}        B. Vakili, H.R. Sepangi, Phys. Lett. B \textbf{651} (2007) 79, [arXiv:0706.0273].
\bibitem{35}        M.V. Battisti, G. Montani, Phys. Lett. B \textbf{656} (2007) 96, [arXiv:gr-qc/0703025].
\bibitem{36}        K. Nozari, and T. Azizi, Gen. Relativ. Gravit. \textbf{38} (2006) 735.
\bibitem{37}        P. Pedram, Europhys. Lett., \textbf{89} (2010) 50008.
\bibitem{38}        P. Wang, H. Yang and X. Zhang, JHEP {\bf08} (2010) 043, [arXiv:1006.5362].
\bibitem{39}        K. Nozari and P. Pedram, Europhys. Lett. \textbf{92} (2010) 50013.
\bibitem{40}        P. Pedram, K. Nozari, and S. H. Taheri, JHEP  {\bf 1103} (2011) 093, [arXiv:1103.1015].
\bibitem{p}         P. Pedram, Int. J. Mod. Phys. D {\bf19} (2010) 2003.
\bibitem{PRD}       P.~Pedram, Phys.~Rev.~D \textbf{85}, 024016 (2012), [arXiv:1112.2327].
\bibitem{PhysA}     P.~Pedram, Physica A \textbf{391}, 2100 (2011).
\bibitem{PLB}       P.~Pedram, Phys.~Lett.~B \textbf{702}, 295 (2011).
\bibitem{9}         S. Das and E.C. Vagenas, Phys. Rev. Lett. {\bf101} (2008) 221301.
\bibitem{14}        L.D. Landau and E.M. Lifshitz, \emph{Quantum Mechanics (Non-relativistic Theory)}, Pergamon Press, Oxford, 1977.
\bibitem{r1}        P. Kennedy, J. Phys. A \textbf{35} (2002) 689.
\bibitem{r2}        A. Arda, O. Aydogdu and R. Sever, J. Phys. A \textbf{43} (2010) 425204.
\bibitem{r3}        C. Rojas and V.M. Villalba, Phys. Rev. A \textbf{71} (2005) 052101.
\end{thebibliography}
\end{document}